\documentclass[final,5p,times,twocolumn]{elsarticle}
\usepackage{verbatim}
\usepackage{listings}
\usepackage{multirow}
\usepackage{graphicx} 
\usepackage{lineno}

\begin{document}
\newcommand{\diana}{{\bf Diana }}
\newcommand{\apollo}{{\bf Apollo }}
\newcommand{\Te}{\ensuremath{\rm^{120}Te} }
\newcommand{\Sn}{\ensuremath{\rm^{120}Sn} }
\newcommand{\TeO}{TeO$_2$ }
\newcommand{\onubb}{$0\nu\beta\beta$}
\newcommand{\onubbplus}{$0\nu\beta^+\beta^+$ }
\newcommand{\onubplusEC}{$0\nu\beta^+$/EC }
\newcommand{\onuECEC}{$0\nu$EC/EC }
\newcommand{\kgy}{kg$\cdot$y }
\newcommand{\bbminus}{$\beta^-\beta^-$ }
\newcommand{\bbplus}{$\beta^+\beta^+$ }
\newcommand{\bplusEC}{$\beta^+$/EC }
\newcommand{\ECEC}{EC/EC }
\newcommand{\Qvalue}{1714.8 $\pm$ 1.3 keV }
\newenvironment{shell}{\vspace{5mm}\newline}{\vspace{5mm}\newline}
\newenvironment{code}{\vspace{5mm}\footnotesize\verbatim}{\endverbatim\normalsize\vspace{5mm}} 
\title{Search for $\beta^+$/EC double beta decay of \Te}

\author[Como]{E.~Andreotti}
\author[Milano,INFNMilano]{C.~Arnaboldi}
\author[USC]{F.~T.~Avignone~III}
\author[LNGS]{M.~Balata}
\author[USC]{I.~Bandac}
\author[Firenze]{M.~Barucci}
\author[LBNL]{J.~W.~Beeman}
\author[Roma]{F.~Bellini}
\author[Milano,INFNMilano]{C.~Brofferio}
\author[LBNL,BerkeleyPhys]{A.~Bryant}
\author[LNGS]{C.~Bucci}
\author[Genova,INFNGenova]{L.~Canonica}
\author[Milano,INFNMilano]{S.~Capelli}
\author[INFNMilano]{L.~Carbone}
\author[Milano,INFNMilano]{M.~Carrettoni}
\author[Milano,INFNMilano]{M.~Clemenza}
\author[INFNMilano]{O.~Cremonesi}
\author[USC]{R.~J.~Creswick}
\author[Genova,INFNGenova]{S.~Di~Domizio}
\author[LLNL,BerkeleyPhys]{M.~J.~Dolinski}
\author[Wisc]{L.~Ejzak}
\author[Roma]{R.~Faccini}
\author[USC]{H.~A.~Farach}
\author[Milano,INFNMilano]{E.~Ferri}
\author[Milano,INFNMilano]{E.~Fiorini}
\author[Como]{L.~Foggetta}
\author[INFNMilano]{A.~Giachero}
\author[Milano,INFNMilano]{L.~Gironi}
\author[Como]{A.~Giuliani}
\author[LNGS]{P.~Gorla}
\author[LNGS,INFNGenova]{E.~Guardincerri}
\author[CalPoly]{T.~D.~Gutierrez}
\author[LBNL,BerkeleyMat]{E.~E.~Haller}
\author[LLNL]{K.~Kazkaz}
\author[Milano,INFNMilano]{S.~Kraft}
\author[LBNL,BerkeleyPhys]{L.~Kogler}
\author[Milano,INFNMilano]{C.~Maiano}
\author[Wisc]{R.~H.~Maruyama}
\author[USC]{C.~Martinez}
\author[INFNMilano]{M.~Martinez}
\author[USC]{S.~Newman}
\author[LNGS]{S.~Nisi}
\author[Como]{C.~Nones}
\author[LLNL,BerkeleyNuc]{E.~B.~Norman}
\author[Milano,INFNMilano]{A.~Nucciotti}
\author[Roma]{F.~Orio}
\author[Genova,INFNGenova]{M.~Pallavicini}
\author[Legnaro]{V.~Palmieri}
\author[Milano,INFNMilano]{L.~Pattavina}
\author[Milano,INFNMilano]{M.~Pavan}
\author[LLNL]{M.~Pedretti}
\author[INFNMilano]{G.~Pessina}
\author[INFNMilano]{S.~Pirro}
\author[INFNMilano]{E.~Previtali}
\author[Firenze]{L.~Risegari}
\author[USC]{C.~Rosenfeld}
\author[Como]{C.~Rusconi}
\author[Como]{C.~Salvioni}
\author[Wisc]{S.~Sangiorgio}
\author[Milano,INFNMilano]{D.~Schaeffer}
\author[LLNL]{N.~D.~Scielzo}
\author[Milano,INFNMilano]{M.~Sisti}
\author[LBNL]{A.~R.~Smith}
\author[Roma]{C.~Tomei}
\author[Firenze]{G.~Ventura}
\author[Roma]{M.~Vignati}

  \address[Como]{Dip.\ di Fisica e Matematica dell'Univ.\ dell'Insubria and Sez.\ INFN di Milano, Como I-22100 - Italy}
  \address[Milano]{Dip.\ di Fisica dell'Universit\`{a} di Milano-Bicocca I-20126 - Italy}
  \address[INFNMilano]{Sez.\ INFN di Mi-Bicocca, Milano I-20126 -Italy}
  \address[USC]{Dept.\ of Phys.\ and Astron., Univ.\ of South Carolina, Columbia, SC 29208 - USA}
  \address[LNGS]{Laboratori Nazionali del Gran Sasso, I-67010, Assergi (L'Aquila) - Italy}
  \address[Firenze]{Dip.\ di Fisica dell'Universit\`{a} di Firenze and Sez.\ INFN di Firenze, Firenze I-50125 - Italy}
  \address[LBNL]{Lawrence Berkeley National Laboratory, Berkeley, CA 94720 - USA}
  \address[Roma]{Dip.\ di Fisica dell'Universit\`{a} di Roma La Sapienza and Sez.\ INFN di Roma, Roma  I-00185 - Italy}
  \address[BerkeleyPhys]{Dept.\ of Physics, Univ.\ of California, Berkeley, CA 94720 - USA}
  \address[Genova]{Dip.\ di Fisica dell'Universit\`{a} di Genova - Italy}
  \address[INFNGenova]{Sez.\ INFN di Genova, Genova I-16146 - Italy}
  \address[LLNL]{Lawrence Livermore National Laboratory, Livermore, CA 94550 - USA}
  \address[Wisc]{Univ.\ of Wisconsin, Madison, WI - USA}
  \address[CalPoly]{California Polytechnic State Univ., San Luis Obispo, CA 93407 - USA}
  \address[BerkeleyMat]{Dept.\ of Materials Sc.\ and Engin., Univ.\ of California, Berkeley, CA 94720 - USA}
  \address[BerkeleyNuc]{Dept.\ of Nuclear Engineering, Univ.\ of California, Berkeley, CA 94720 - USA}
  \address[Legnaro]{Laboratori Nazionali di Legnaro, I-35020 Legnaro (Padova) - Italy}

\begin{abstract}
We present a search for $\beta^+$/EC double beta decay of \Te performed with the CUORICINO experiment,
an array of $\rm{TeO}_2$ cryogenic bolometers. After collecting 0.0573~\kgy of $^{120}\rm{Te}$,  we see no evidence of a signal and therefore set the following limits on the half-life: $\rm T^{0\nu}_{1/2}> 1.9 \cdot 10^{21}$~y at 90\% C.L. for the $0\nu$ mode and $\rm T^{2\nu}_{1/2}> 7.6 \cdot 10^{19}$~y at 90\% C.L. for the $2\nu$ mode. These results improve the existing limits by almost three orders of magnitude (four in the case of $0\nu$ mode). 
\end{abstract}
\maketitle

The discovery of neutrino oscillations~\cite{oscillation} proved that neutrinos are massive, but there are fundamental issues that oscillation experiments cannot address: measuring the absolute neutrino mass and determining whether the neutrino is the antiparticle of itself, thus being of Majorana nature, or not.
To answer these questions it is necessary to look for  neutrinoless double beta, \onubb, decays, which would bring conclusive evidence of the Majorana nature of the neutrino and whose decay rates constrain the absolute neutrino mass~\cite{dbeta}. 

Double beta decays can occur by either emitting two electrons or two positrons. In the latter case, either of the positron emissions can be replaced by an electron capture (EC). While $\beta^-\beta^-$ decays have the largest expected rates, $\beta^+/EC$  and $\beta^+\beta^+$decays provide clear signatures from the 511-keV annihilation gamma rays. Energy and momentum conservation in the EC/EC decay requires an extra radiative process, reducing the rate by several orders of magnitude.

This paper reports on a search for $\beta^+$/EC  decays $\Te\to\Sn + e^+$ and $\Te\to\Sn + e^+ + 2\nu$ with the CUORICINO experiment, an array of \TeO cryogenic bolometers at the Gran Sasso National Laboratories. 
The  \Te isotope has been only minimally investigated from a theoretical point of view. No calculations of the half-life of \onubplusEC decay of \Te are available for comparison with our result. The predictions for the same decay mechanism in other nuclei, assuming the effective Majorana mass $\langle m_\nu \rangle$ = 1~eV, range between  $10^{26}$--$10^{27}$~y~\cite{theo,theo2}. The larger value is mostly due to the need to have simultaneously a decay and a capture and to the reduced phase space available. 
The only reference for the two neutrino mode~\cite{Abad} yields a theoretical value for the half-life of  $4.4 \cdot 10^{26}$~y. 

In recent years, experimental limits on the \Te decays have been set using an array of CdZnTe detectors located at the Gran Sasso Underground Laboratory~\cite{Cobra07,Cobra09} and a HPGe detector at the Modane Underground Laboratory~\cite{Bar07,Bar08}. The present best limits in the literature are T$^{0\nu}_{1/2} > 4.1 \cdot 10^{17}$~y~\cite{Cobra09} and T$^{(0\nu+2\nu)}_{1/2} > 1.9 \cdot 10^{17}$~y~\cite{Bar07}.

The search strategy is the following: the emitted positron carries a kinetic energy up to K$_{max}$ = Q - 2m$_e$c$^2$ - E$_b$, where $E_b$ is the binding energy of the captured electron within the atomic shell and Q$=(1714.8\pm1.3)$~keV~\cite{Scielzo} is the difference in $^{120}$Te and $^{120}$Sn atomic masses. The electron capture is most likely to occur from the K shell, whose binding energy is 30.5 keV. The ratio of L-capture to K-capture for most elements is around 10\% (12\% for$^{120}$Sb $\rightarrow$ $^{120}$Sn EC decay)~\cite{EC}. In the following we will always assume a capture from the K shell.

The bolometer where the decay occurs will see the deposition of both the binding energy and the kinetic energy of the positron (maximum energy: $E_{0}=K_{max}+E_b=Q-2m_ec^2=692.8$~keV, independent of $E_b$).\footnote{Here and in the following we assume that the X-rays following the EC do not escape from the crystal where the decay occurs.} Once at rest, the positron annihilates with an electron and two photons with an energy $E_\gamma=511.0$~keV are emitted. These photons can interact with the same bolometer or by a nearby one, or they can escape undetected. 

The analysis presented here searches for the signatures with the best signal-to-noise ratio. This is not the case for the events where the entire energy is deposited inside the detector, due to the low detection efficiency (see Table~\ref{Tab1}). For the $0\nu$ mode, where the positron is monochromatic, this means the coincidence between a bolometer with an energy deposition consistent with $E_\gamma$ and one with either $E_0$ or $E_0+E_\gamma$ and the coincidence of one bolometer with a signal of $E_0$ and two other bolometers with a signal of $E_\gamma$. For the $2\nu$ mode, where the positron is emitted with a continuum of kinetic energies between 0 and K$_{max}$, this means the triple coincidence of one bolometer with a signal between $E_b$ and $E_0$ and two others with a signal of $E_\gamma$. 

\section{The CUORICINO detector}
\label{sec:qino}
\begin{figure}[t]
\centering
\includegraphics[width=30mm]{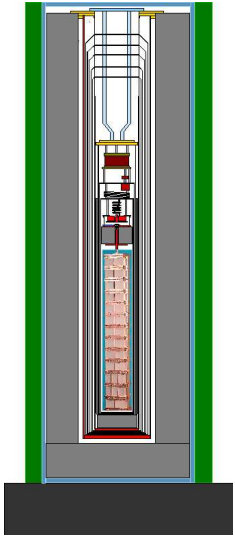}
\caption{A sketch of the CUORICINO assembly showing the tower hanging from the mixing chamber, the various heat shields and the external shielding.}\label{Fig1}
\end{figure}

The CUORICINO experiment is detailed in Ref.~\cite{Qino08}. Briefly, it is an array of \TeO crystals acting as cryogenic bolometers at a working temperature of 8--10 mK and with heat capacity $~2.3\cdot10^ {-9}$~J/K.
To measure temperature variations corresponding to few keV ($\Delta T\sim 0.1~\mu$K/keV) heavily doped high-resistance germanium thermistors (NTD, Neutron Transmutation Doped) are glued to each crystal. 

The CUORICINO detector consists of 62 \TeO crystals arranged in 13 planes.  
Each of the upper 10 planes and the lowest one consists of four $5\times5\times5$~cm$^3$ \TeO crystals, while the 11th and 12th planes have nine, $3\times3\times6$~cm$^3$ crystals. 
All crystals have natural isotopic abundances except four of the smaller crystals, two of which are enriched to 82.3\% in $^{128}$Te and two to 75\% in $^{130}$Te. 

The natural abundance of \Te is 0.096\%~\cite{TOI}, so the 39.4~kg of the CUORICINO experiment (enriched crystals are not included) contain $N_{\beta\beta}=1.43 \cdot 10^{23}$~nuclei of $^{120}$Te . 

The experiment is shielded with two layers of lead of 10~cm minimum thickness each. The outer layer is made of common low radioactivity lead, while the inner layer is made of special lead with a low activity of $^{210}$Pb. The electrolytic copper of the refrigerator thermal shields provides an additional shield with a minimum thickness of 2~cm. An external 10~cm layer of borated polyethylene was installed to reduce the background due to environmental neutrons. The detector itself is shielded against the intrinsic radioactive contamination of the dilution unit materials by an internal layer of 10~cm of Roman lead~\cite{Qino98}, located inside the cryostat immediately above the tower. The background from the activity in the lateral thermal shields of the dilution 
refrigerator is reduced by a lateral internal 1.4~cm thick shield of Roman lead. Another 8~cm lead shield is located at the bottom of the tower. The refrigerator is surrounded by a Plexiglas anti-radon box flushed with clean $N_2$ from a liquid nitrogen evaporator and is also enclosed in a Faraday cage to eliminate electromagnetic interference. A sketch of the assembly is shown in Fig.~\ref{Fig1}. 
\begin{table}[h]
\begin{center}
\begin{tabular}{|c|c|c|} \hline
Signature [energies in keV] &$\mu$ & $\varepsilon$ [\%] \\ 
\hline \hline
 (30.5 -- 692.8)                                        & 1 &  3.00 $\pm$ 0.02 \\ \hline 
(30.5 -- 692.8) + 511                    & 2 &  3.40 $\pm$ 0.02 \\ \hline
(30.5 -- 692.8) + 511 + 511  & 3 & 0.45 $\pm$ 0.01\\ \hline
(541.5 -- 1203.8)                                    & 1 & 16.28 $\pm$ 0.04\\ \hline 
(541.5 -- 1203.8) + 511              & 2 & 6.23 $\pm$ 0.03\\ \hline
(1052.5 -- 1714.8)                                 & 1 & 10.04 $\pm$ 0.03\\ \hline
\end{tabular}
\caption{Signatures of \Te \bplusEC decay in an array of \TeO detectors and their corresponding multiplicity ($\mu$), that is the number of detectors with an energy deposition above threshold. The detection efficiency for the $0\nu$ mode in CUORICINO is reported in the last column ($\varepsilon$). We denote with the $+$ sign the coincidence of energies released in different detectors. For the $0\nu$ mode the energy released in the detector where the decay occurred corresponds to the upper bound of the interval. The errors are statistical only.}
\label{Tab1}
\end{center}
\end{table}
\begin{figure*}[htb]
\centering
\includegraphics[width=0.9\linewidth]{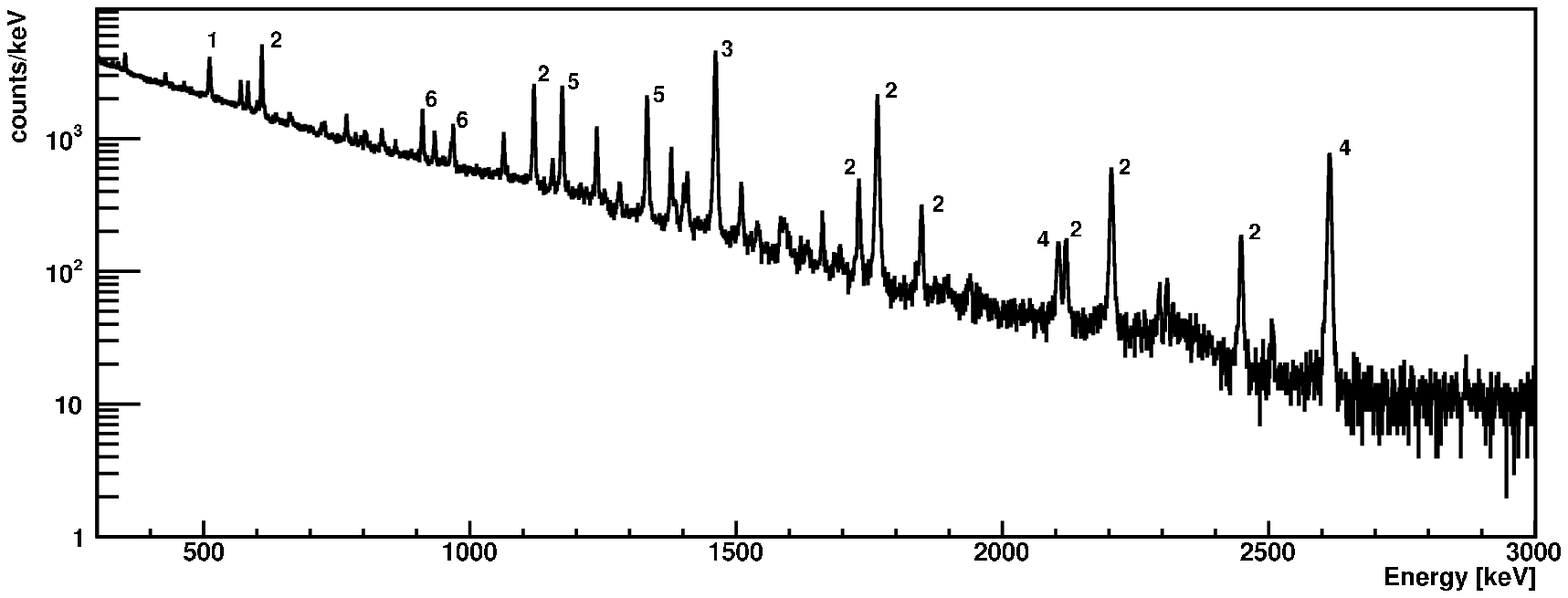}
\caption{Total energy spectrum of all CUORICINO detectors. The most prominent peaks are labeled and come from known radioactive sources such as: $e^+e^-$ annihilation (1), $^{214}$Bi (2), $^{40}$K (3), $^{208}$Tl (4), $^{60}$Co (5) and  $^{228}$Ac (6).}\label{Fig2}
\end{figure*}

\begin{figure*}[htb]
\centering
\includegraphics[width=0.45\linewidth]{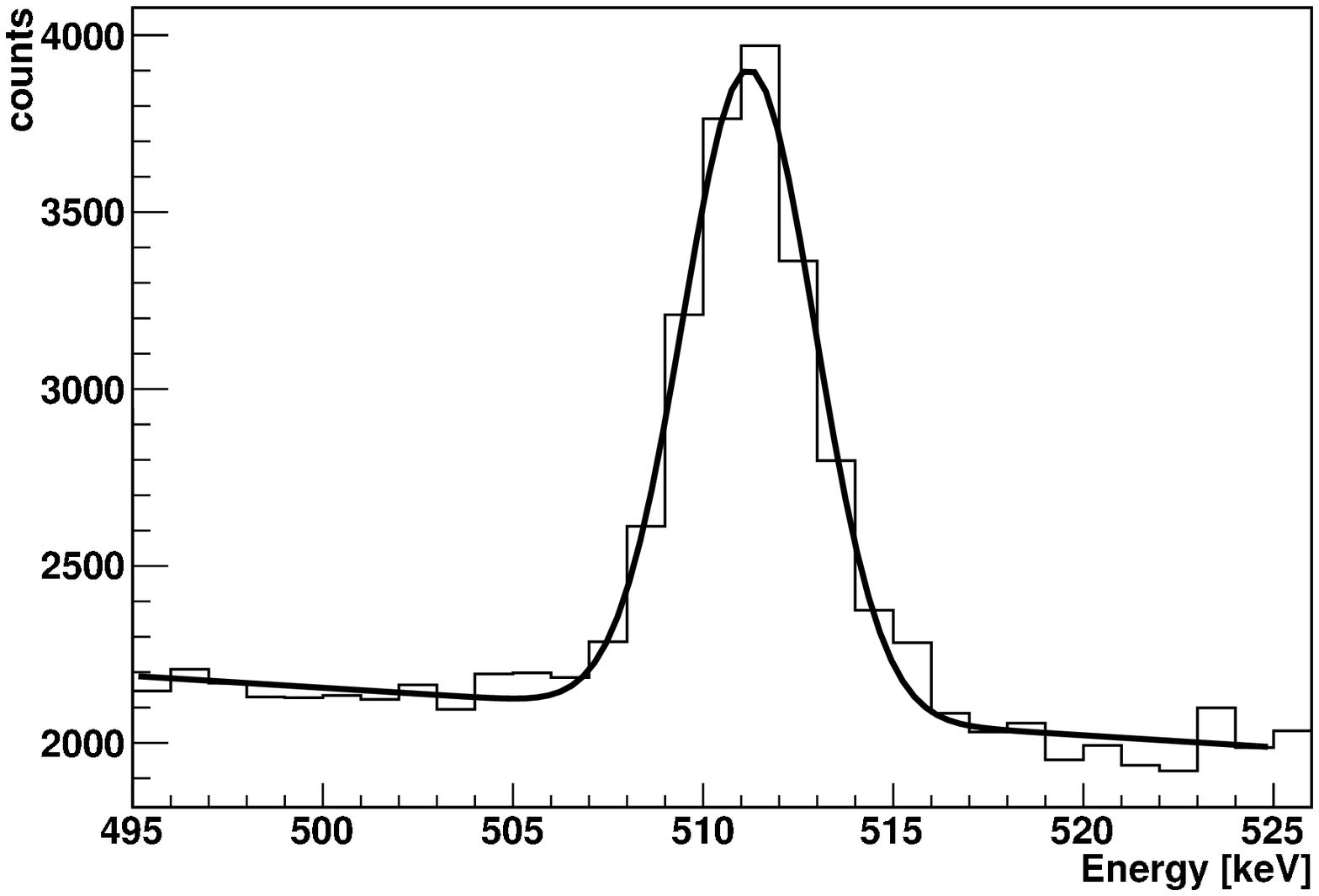}
\includegraphics[width=0.45\linewidth]{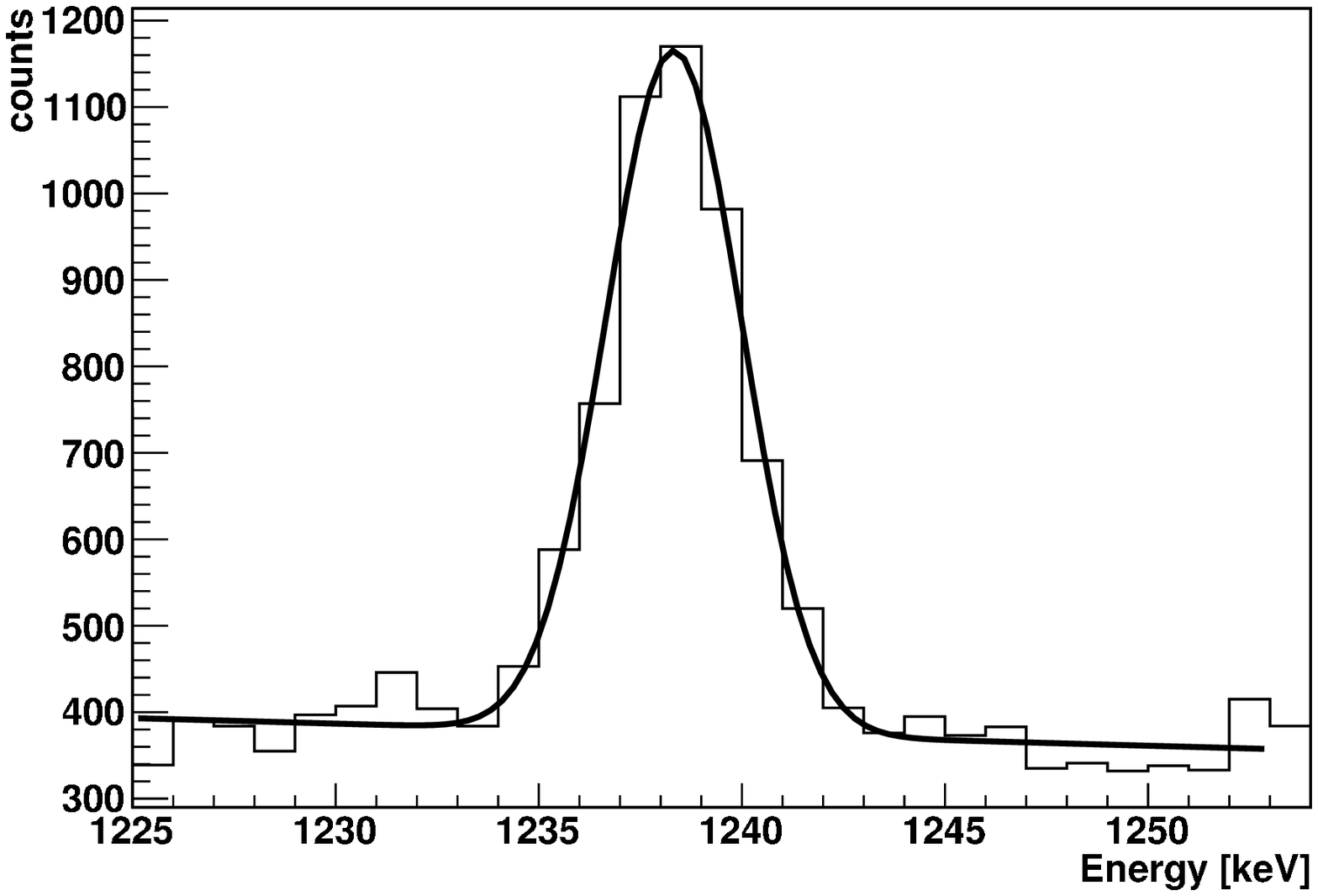}
\caption{Resolution fits at E=511~keV (left) and E=1238~keV(right). The fit function is a Gaussian plus linear background.}\label{Fig3}
\end{figure*}

For the present analysis, the full CUORICINO statistics (data collected between May 2004 and May 2008) for a total exposure of 0.0573~\kgy of $^{120}\rm{Te}$ is used. The total energy spectrum of all detectors is shown in Fig.~\ref{Fig2}. 
Several peaks of radioactive isotopes are clearly visible, the most prominent are labeled.
\section{Data acquisition and analysis}
CUORICINO data are divided in datasets, each one being a collection of about a month of daily measurements. Routine calibrations are performed at the beginning and at the end of each dataset using two wires of thoriated tungsten inserted inside the external lead shield.
The signals coming from each bolometer are amplified and filtered with a six-pole Bessel low-pass filter and fed to a 16-bit ADC. The signal is digitized with a sampling time of 8~ms. With each triggered pulse, a set of 512 samples is recorded to disk.
The typical bandwidth is approximately 10~Hz, with signal rise and decay times of order 30 and 500~ms, respectively. More details of the design and features of the electronics system are found in Ref.~\cite{Arn02}. 

Each bolometer has a different trigger threshold, optimized according to the bolometer's typical noise and pulse shape. The trigger rate is time and channel dependent, with a mean value of about 1~mHz. 
The amplitude of the pulses is estimated by means of an Optimal Filter technique~\cite{Qino04}. 
The gain of each bolometer is monitored by means of a Si resistor of 50--100~k$\Omega$ attached to it that acts as a heater. Heat pulses are periodically supplied by an ultra-stable pulser~\cite{Arn03} that sends a calibrated voltage pulse to the Si resistors. Their Joule dissipation produces heat pulses in the crystal with a shape which is almost identical to the calibration $\gamma$-rays. 
%
\begin{figure*}[htb]
\centering
\includegraphics[width=0.45\linewidth]{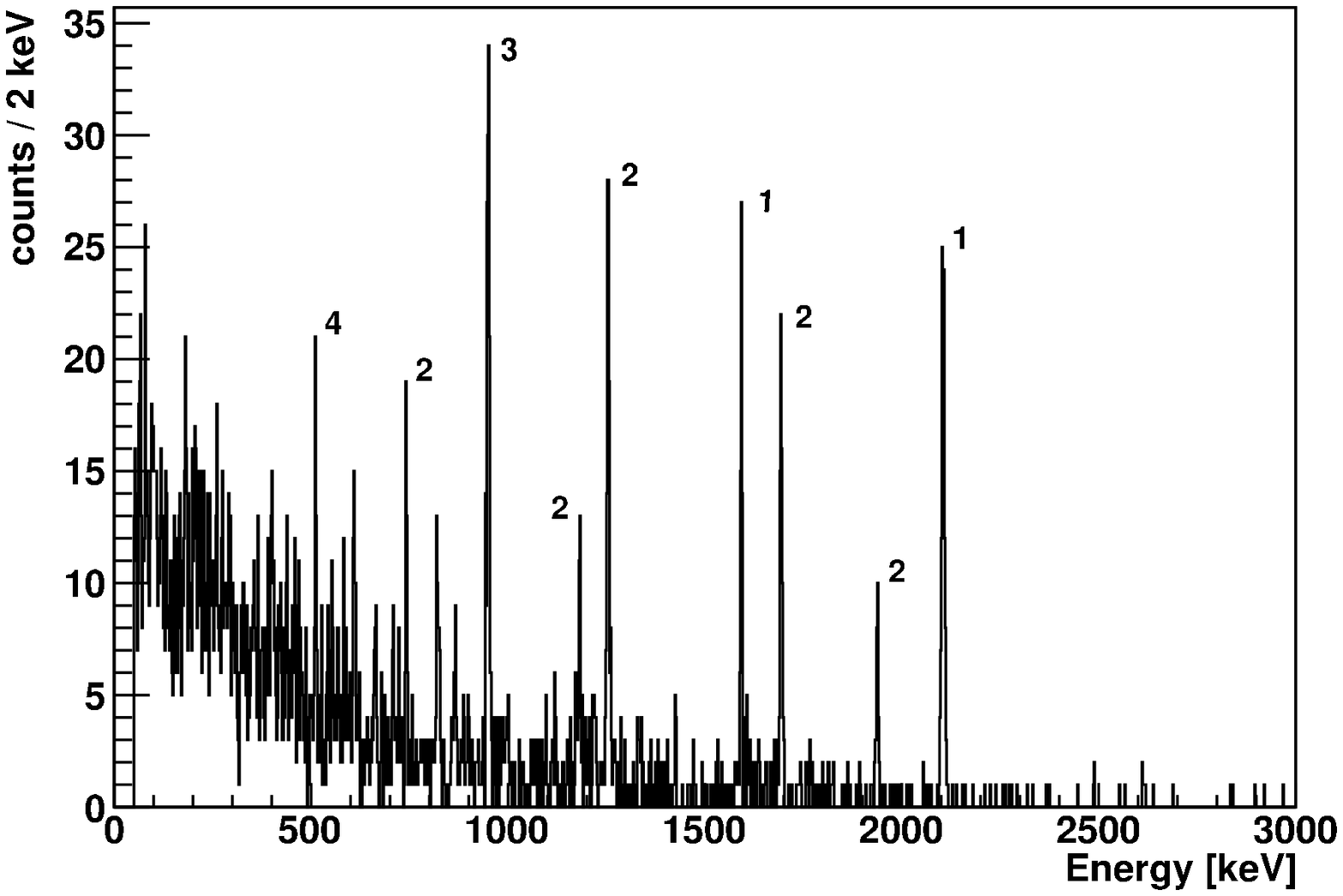}
\includegraphics[width=0.45\linewidth]{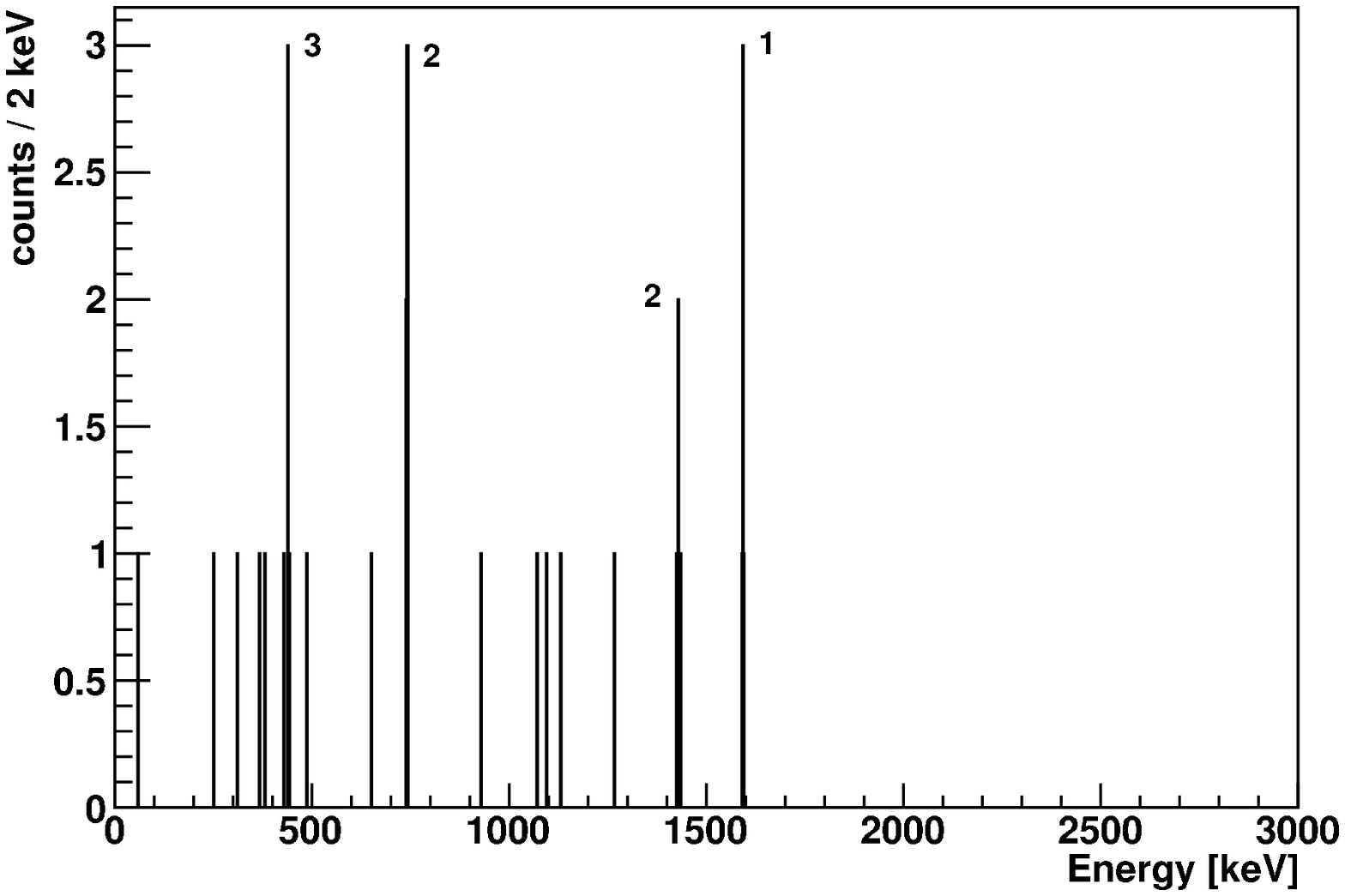}
\caption{Energy spectra of coincidences with a single 511~keV event (left) and of coincidences with two 511~keV photons (right). The most prominent peaks are labeled and come from the single escape and the double escape of known lines from the following isotopes: $^{208}$Tl (1), $^{214}$Bi (2) and $^{40}$K (3). In the spectrum of double coincidences (left) the 511~keV  peak is also visible (4).}\label{Fig4}
\end{figure*}
\begin{figure}[h]
\centering
\includegraphics[width=.8\linewidth]{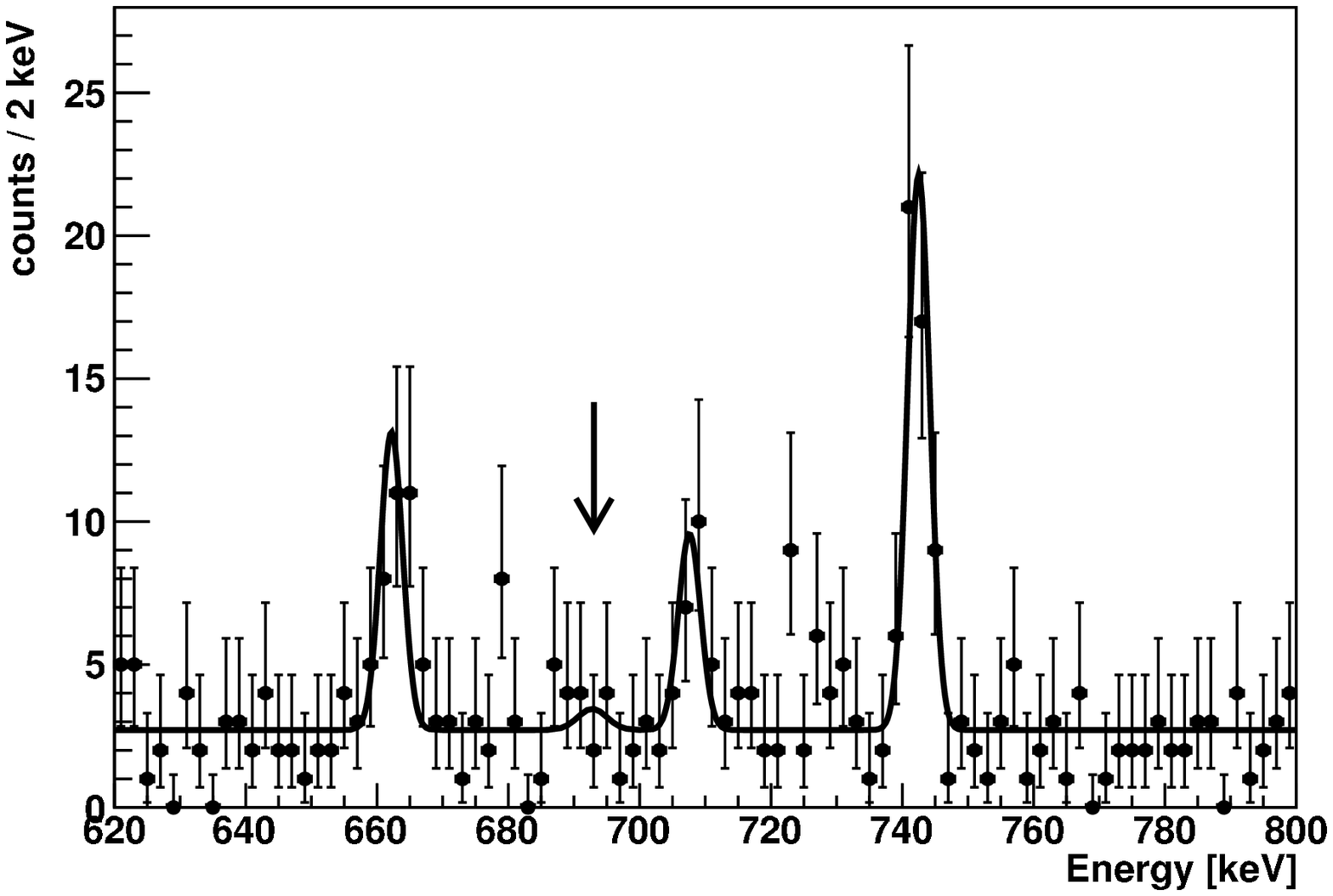}
\includegraphics[width=.8\linewidth]{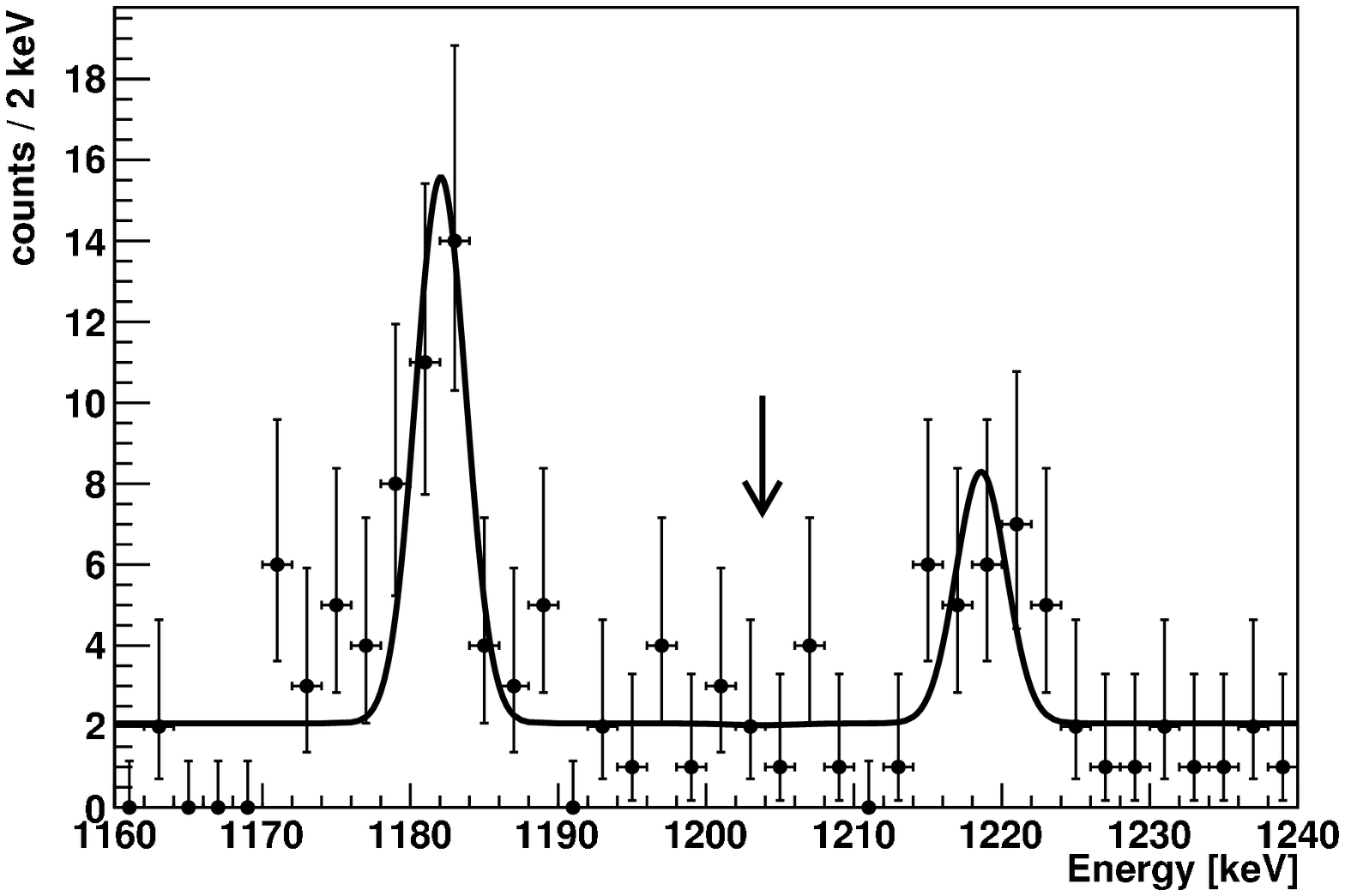}
\includegraphics[width=.8\linewidth]{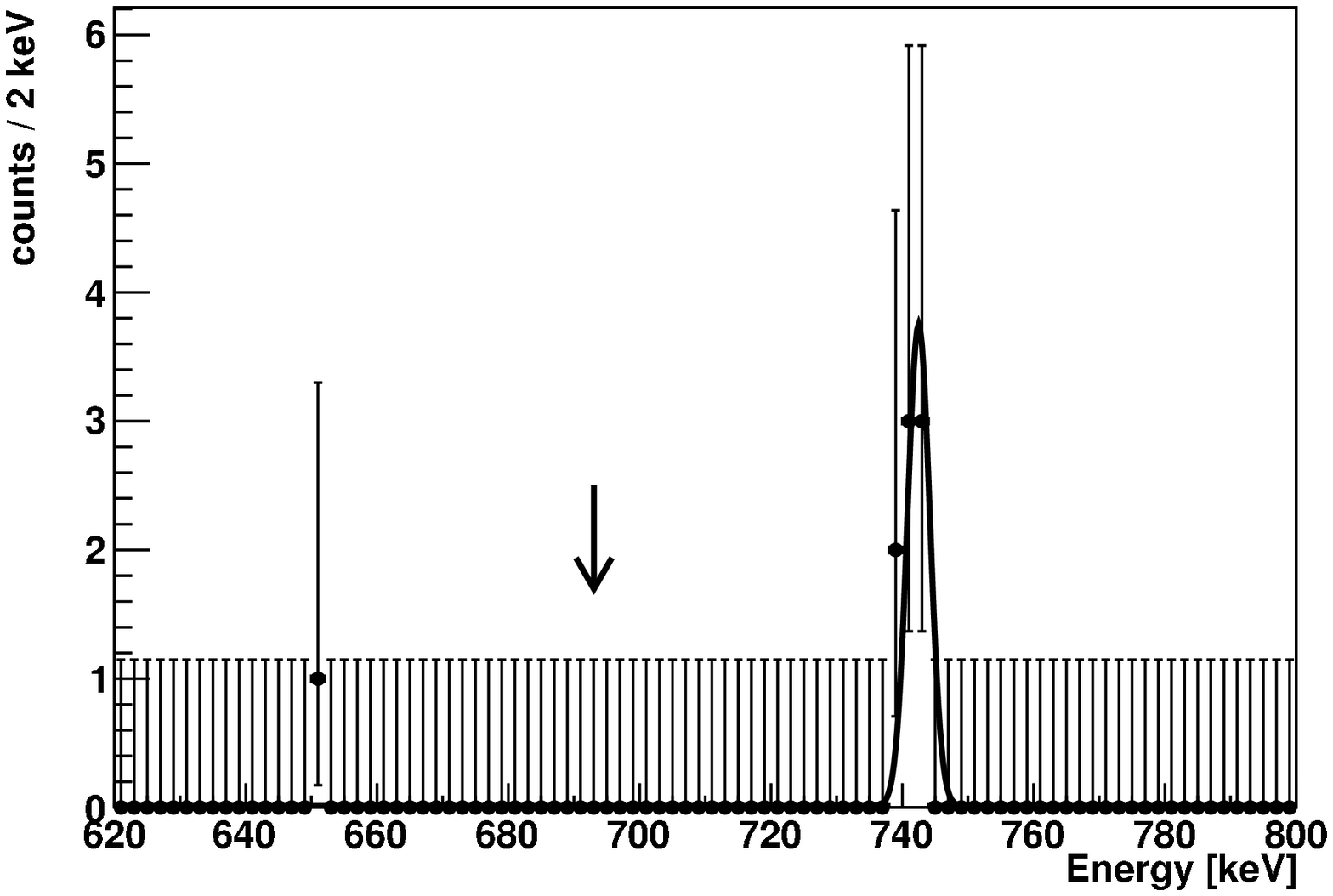}
\caption{Energy spectra, in the regions of interest, of coincidences with a single 511~keV event [two instances] (top) and of coincidences with two 511~keV photons (bottom). The arrows indicate the energy of the expected signal. Fit results, as explained in the text, are overlaid.}\label{Fig5}
\end{figure}

\section{Search strategy}

As detailed in the introduction, a \bplusEC decay of \Te in an array of \TeO detectors releases an energy up to $E_0=692.8$~keV in the bolometer where the decay occurs and, if the annihilation gammas do not escape undetected, one or two additional energy deposits of $E_\gamma=511.0$~keV in either the same or a nearby bolometer. There are therefore several distinctive signatures as listed in Tab.~\ref{Tab1}.

For the $0\nu$ mode the energy released in the detector where the decay occurred corresponds to the upper bound of the interval. 
The detection efficiencies for the $0\nu$ mode were estimated by means of a GEANT4 simulation of the CUORICINO setup (see last column of Tab.~\ref{Tab1}) where the decays are located uniformly within all non-enriched detectors and the decay products are emitted isotropically. We always assume that the binding energy of the captured electron is released within the detector where the decay occurs. 
The efficiency estimate includes the dead time evaluated separately for each detector. To account for the different energy resolution of CUORICINO detectors we assigned to each detector in the simulation a FWHM given by the weighted mean of the FWHMs calculated in all CUORICINO calibration runs.
The highest detection efficiencies correspond to the cases where one or both of the electron-positron annihilation photons are fully absorbed in the detector where the \bplusEC decay occurs. This is consistent with the relatively large size of the CUORICINO detectors and the mean free path of a 511~keV photon in \TeO (1.9~cm). 
These signatures are calculated to account for nearly half of all decays.  The remainder involve only partial energy deposition of the 511~keV gamma ray energy and therefore are more difficult to distinguish from background.
Due to the extremely short range of positrons in CUORICINO crystals, the efficiencies estimated for the $0\nu$ mode are also valid for the $2\nu$ mode, apart from small corrections (see below).
Considering also the expected background, we limited our analysis to the signatures that feature a coincidence of two or three events for the $0\nu$ mode and of three events for the $2\nu$ mode.

We consequently search only for double or triple coincidences where one or two of the energy depositions are in the interval $\pm2.5\sigma$ of $E_\gamma=511$~keV, where $\sigma=$ 1.7~keV, as estimated in the inclusive energy spectrum (see Fig.~\ref{Fig3}). The coincidence window is 100~ms. The probability of accidental coincidence is estimated  from the measurement of the single crystal rate ($\sim$ 1.6~mHz) to be around 0.7\% for double coincidences between detectors in the same plane or in adjacent planes. For triple coincidences it is even less.

The energy spectra of the events in coincidence with one or two 511~keV photons are shown in Fig.~\ref{Fig4}. The structures in both spectra have been identified and correspond to single or double escape lines from known radioactive lines 
present in the CUORICINO background spectrum.
The continuum background is due both to accidental coincidences and true coincidences in which the energy deposition of the event that accompanies the 511 keV gamma(s) is not complete.

\section{Results of the 0$\nu\beta^+$/EC decay search}

In the spectra of Fig.~\ref{Fig4} we search for a signal with mean energy 692.8~keV (spectra of double and triple coincidences) or 1203.8~keV (spectrum of double coincidences only). The expected resolution at 1203.8~keV is estimated on the $^{214}$Bi line at 1238~keV and found to be $\sigma$ = 1.67 $\pm$ 0.05~keV (see Fig.~\ref{Fig3}), in agreement with the value 
of 1.7~keV found for the 511 keV~line and consistent with being constant over the energy region of interest.

Fig.~\ref{Fig5} shows the energy regions of interest of the double and triple coincidences spectra. The observed lines, as explained above, correspond to single- and double-escape peaks. Specifically, the measured spectra show the single escape lines of the 1173.2~keV transition from $^{60}$Co at 662.2~keV and of the 1729.6~keV transition from $^{214}$Bi at 1218.6~keV plus the double escape lines of the 2204.06~keV transition from $^{214}$Bi at 1182.06~keV, of  the 1764.5~keV transition from $^{214}$Bi at 742.5~keV and again of the 1729.6~keV transition from $^{214}$Bi at 707.6~keV. The observed SE and DE peaks produce a negligible contribution to the background in the energy regions of interest.

The search for the  0$\nu$$\beta^+$/EC decay of $^{120}$Te on the spectra in Fig.~\ref{Fig5} is performed with unbinned maximum likelihood fits. 
In the likelihood fit the background is therefore parametrized with the sum of a flat component and a Gaussian for each of  the expected escape lines, with mean values fixed to the corresponding known energies and the width fixed to $\sigma=1.7$~keV. The signal is also parametrized with a Gaussian with mean value fixed to the expectations, that is 692.8~keV or 1203.8~keV, depending on whether or not one of the annihilation photons is absorbed in the same crystal where it is emitted. The width of the signal line is fixed to 2.2~keV which is obtained by summing in quadrature the sigma (1.7~keV) , the error on the Q-value (1.3~keV), and the uncertainty on the energy scale (0.4~keV) as estimated from the calibration fit\footnote{From a bayesian point of view (i.e. our approach), you should sum the likelihoods for different Q-values weighting them by the probability of that Q-value being correct. This is equivalent to convoluting the peak position with a gaussian, and therefore to a gaussian whose width is the sum in quadrature of the uncertainty on the Q-value and the resolution.}. In summary for each spectrum the fitted quantities are the number of events in the signal, the flat background and the escape lines. The curves resulting from the fits are overlaid in Fig.~\ref{Fig5} and the corresponding fitted quantities are reported in Tab.~\ref{tab:results}.

The upper limit on the number of observed signal events is extracted by means of a Bayesian procedure.
By integrating with a Monte Carlo technique the likelihood over the nuisance parameters, we calculate, for each signature, the posterior probability density function ({\it pdf}) for the parameter $N_{sig}$ and define our 90\% C.L. upper limit as the value where its integral reaches the 90\% of the total area (see Tab.~\ref{tab:results}).

To combine the results coming from the three signatures, we calculate with the same procedure the pdfs for the parameter $N_{sig}/\epsilon_{tot}$, the number of signal events divided by the corresponding efficiency. The efficiency is calculated as $\epsilon_{tot}=\epsilon\times(\epsilon_{noise}\epsilon_{heat})^{\mu}$, where  $\epsilon$ is the efficiency tabulated in Tab.~\ref{Tab1}, $\mu=2,3$ is the multiplicity of the given signature from the same table, $\epsilon_{noise}=99.1\%$ accounts for the loss of signal due to noise as estimated on heater pulses, and  $\epsilon_{heat}=97.7\%$ accounts for the dead time induced by the presence of heater. This procedure, together with the inclusion of the uncertainty on the energy scale in the fits of the spectra, accounts for the systematic errors, which have a negligible impact on the result.

From the combined pdf, we extract a 90\% C.L. upper limit on the total number of double beta decays regardless of their detection, $n_B=100$.
We can then set a limit on the half-life of 0$\nu\beta^+$/EC decay of $^{120}$Te
\begin{equation}
T^{0\nu}_{1/2} > \ln2N_{\beta\beta} \frac{T}{n_{B}}=1.9 \cdot 10^{21}~y 
\end{equation}
where $N_{\beta\beta}$ is the number of $^{120}$Te nuclei and $T$ is the live time.

\section{Results of the 2$\nu\beta^+$/EC decay search}
\begin{figure}[h]
\centering
\includegraphics[width=80mm]{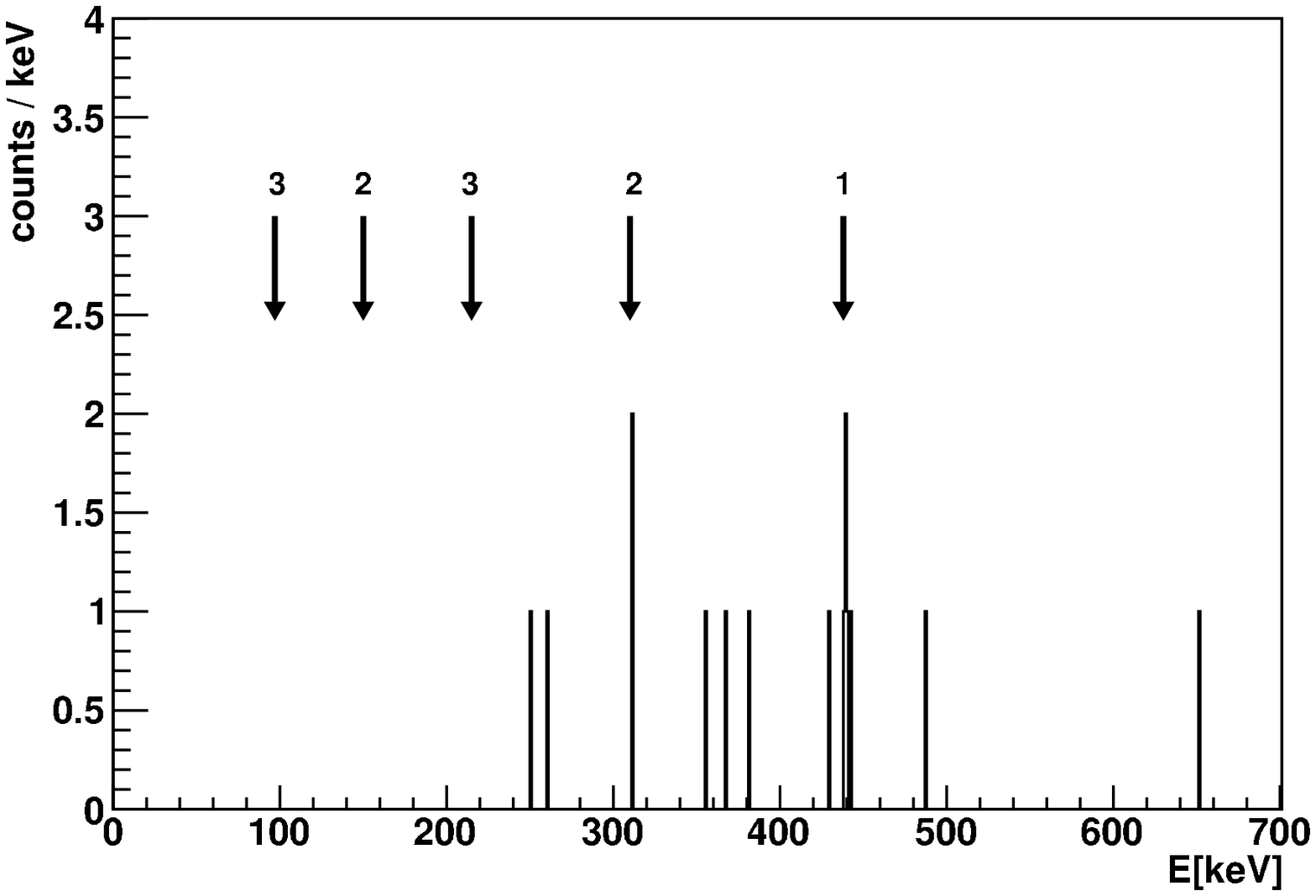}
\caption{Energy spectrum of triple coincidences with two 511~keV photons in the energy region where we expect a signal from the 2$\nu$$\beta^+$/EC. The red arrows indicate the energies where we expect the DE peaks from known radioactive gamma lines in the CUORICINO inclusive spectrum. For each of these energies we subtract an energy region of $\pm 3\sigma$. The lines that can give a double-escape peak between 50 keV and 692.8 keV are 1120.3~keV and 1238.1~keV of $^{214}$Bi (3), 1173.2~keV and 1332.5~keV of $^{60}$Co (2) and 1460.8~keV of $^{40}$K (1).}\label{Fig6}
\end{figure}

The energy spectrum of the events in coincidence with two 511~keV photons in the region below 692.8 keV contains 15 events (see Fig.~\ref{Fig6}). We can exclude from our analysis the regions $\pm 3\sigma$ around the energies where we expect the DE peaks from known radioactive gamma lines in the CUORICINO inclusive spectrum. These energies are indicated with red arrows in Fig.~\ref{Fig6} and listed in the caption. We have listed only the gamma lines that are expected to contribute with at least one event. This expectation is based on the comparison with the $^{40}$K line that produces 5 events at 438.8~keV in the experimental spectrum of triple coincidences (see Fig.~\ref{Fig6}, line labeled as 1). 
After the subtraction, only 8 events remain in the spectrum from threshold to end-point. These events can be due to accidental coincidences or to true coincidences (double escape of already compton scattered gammas). To estimate the first component
we looked at the spectra of events in triple coincidence with the side-band of 511 keV (left side-band: from 470 keV to 502.5 keV, right side-band: from 519.5 keV to 560 keV) correctly normalized to the experimental spectrum of Fig.~\ref{Fig6}. This accounts for $4.3\pm0.5$ events. The other component of the background cannot be reliably estimated due to limited statistics and we cannot discriminate against it.  Therefore, we set an upper limit assuming conservatively that the remaining events may be signal.

\begin{table*}[]
\begin{center}
\begin{tabular}{|| l || c |c|c|c|c|c|} \hline
spectrum&$N_{sig}$ &$N_{flat}$ &$N_{L1}$ & $N_{L2}$ & $N_{H}$ & 90\% U.L. \\ 
\hline \hline
I   &  1.7$\pm$3.5 & 214$\pm$16 & 17.5$\pm$5.3 & 14$\pm$5 & 34$\pm$6 & 8.5\\
II  &  0.3$\pm$3.1 & 78$\pm$10 & 26$\pm$6 & N/A & 9.5$\pm$4.3 & 7.25\\
III &  0.0$\pm$0.5 & 1$\pm$1 & N/A & N/A & 8$\pm$3 & 2.62\\ \hline
\end{tabular}
\caption{Results of the unbinned likelihood fits to the energy spectrum of the double coincidences around $E=692.8$~keV ("I", shown in the top plot of Fig.~\ref{Fig5}),  around $E=1209.8$~keV ("II", shown in the middle plot of Fig.~\ref{Fig5}), and the triple coincidences around $E=692.8$~keV ("III", shown in the bottom plot of Fig.~\ref{Fig5}). Here $N_{sig}$ estimates the number of signal events, $N_{flat}$
the total number of flat background events in the fit interval, $N_{L1}$ and $N_{L2}$ are the number of events in the escape peaks of lower energies and $N_{H}$ is the number of events in the escape peak of higher energies. Finally, the 90\% C.L. upper limits on the number of observed events are reported.}
\label{tab:results}
\end{center}
\end{table*}

The upper limit on the number of observed signal events is extracted by means of a Bayesian procedure.
From the likelihood function for Poisson distributed data with unknown mean $s$ and known background $b$ and using as prior pdf a flat prior different from zero only for positive values of $s$, we extract our 90\% C.L. upper limit on the number of observed events as the value where the integral of the posterior pdf reaches the 90\% of the total area~\cite{PDG}, $n_B=9.04$. We can then set a limit on the half-life of 2$\nu\beta^+$/EC decay of $^{120}$Te
\begin{equation}
T^{2\nu}_{1/2} > \ln2N_{\beta\beta} \frac{\epsilon_{tot}T}{n_{B}}=0.76 \cdot 10^{20}~y 
\end{equation}
where $N_{\beta\beta}$ is the number of $^{120}$Te nuclei and $T$ is the live time.  The efficiency is calculated as $\epsilon_{tot}=\epsilon\times(\epsilon_{noise}\epsilon_{heat})^3\times(1-\epsilon_{corr})$, where $\epsilon$, $\epsilon_{noise}$ and $\epsilon_{heat}$ have been defined above and $\epsilon_{corr}=12.5\pm0.5\%$ is the correction to be applied to the detection efficiency of the $0\nu$ mode and accounts for the fact that we are not sensitive to the portion of the positron spectrum between 30.5 and 50 keV (threshold) and that we subtract five 10 keV energy regions from the experimental spectrum, as explained above. Since we do not know the exact shape of the positron spectrum from the 2$\nu$$\beta^+$/EC decay of \Te the estimation of $\epsilon_{corr}$ has been performed on a standard $\beta$ spectrum with end-point at 692.8 keV and the error is estimated from the comparison of $\epsilon_{corr}$ estimated as explained above and $\epsilon_{corr}$ estimated on a flat spectrum from 0 to end point. Systematic errors on the efficiencies have a negligible impact on the result.

\section{Conclusions}
We  searched for double beta decays $\beta^+$/EC of  \Te in the $\rm{TeO}_2$ cryogenic bolometers of CUORICINO, using  0.0573~\kgy of $^{120}\rm{Te}$ of data.  We see no evidence of a signal and therefore set new limits on the half-life for $0\nu$ and $2\nu$ decay $\rm T^{0\nu}_{1/2}> 1.9 \cdot 10^{21}$~y and $\rm T^{2\nu}_{1/2}> 0.76 \cdot 10^{20}$~y extending the exclusion region by almost three orders of magnitude (four in the case of $0\nu$ mode) with respect to the constraints from previous experiments. 
The limits obtained with CUORICINO could be improved in the future by an additional $\sim$2 orders of magnitude with the CUORE experiment~\cite{CUORE} because of increased mass, higher coincident detection efficiency, and lower backgrounds.


\begin{thebibliography}{99}
\bibitem{oscillation} 
 Y.~Fukuda et al. [Super-Kamiokande Collaboration],
  Phys.\ Rev.\ Lett.\ 81 (1998) 1562,
  Y.~Fukuda et al. [Super-Kamiokande Collaboration],
  Phys.\ Rev.\ Lett.\ 82 (1999) 2430,
  S.~Fukuda et al. [Super-Kamiokande Collaboration],
  Phys.\ Rev.\ Lett.\ 86 (2001) 5651,
  Q.~R.~Ahmad et al. [SNO Collaboration],
  Phys.\ Rev.\ Lett.\ 87 (2001) 071301,
  Q.~R.~Ahmad et al. [SNO Collaboration],
  Phys.\ Rev.\ Lett.\ 89 (2002) 011301,
  K.~Eguchi et al. [KamLAND Collaboration],
  Phys.\ Rev.\ Lett.\ 90 (2003) 021802,
  T.~Araki et al. [KamLAND Collaboration],
  Phys.\ Rev.\ Lett.\ 94 (2005) 081801.
  
\bibitem{dbeta} Examples of recent reviews are 
  S.~Elliott and P.~Vogel, Ann.\ Rev.\ Nucl.\ Part.\ Sci.\ 52 (2002) 115,
  A.~Morales and J.~Morales, Nucl.\ Phys.\ B (Proc. Suppl.) 114 (2003) 141, 
  F.~T.~Avignone III, S.~R. Elliott and J.~Engel (2006), Rev.\ Mod.\ Phys.\ 80 (2008) 481 and
  K.~Zuber, Acta Phys.\ Polon.\  B 37 (2006) 1905.
    
\bibitem{theo} V.I.~Tretyak and Yu.G.~Zdesenko, ATOMIC DATA AND NUCLEAR DATA TABLES 61, 43 (1995). 
\bibitem{theo2} V.I.~Tretyak and Yu.G.~Zdesenko, ATOMIC DATA AND NUCLEAR DATA TABLES 80, 83 (2002). 
\bibitem{Abad} J.~Abad et al., J. de Physique 45 C3 (1984).
\bibitem{Cobra07} T.~Bloxham et al., Phys.\ Rev.\ C 76, (2007) 025501.
\bibitem{Cobra09} J.V.~Dawson et al., Phys.\ Rev.\ C 80, (2009) 025502.
\bibitem{Bar07} A.S.~Barabash et al., J.\ Phys.\ G 34 (2007) 1721-1728.
\bibitem{Bar08} A.S.~Barabash et al., J.\ Phys.\ Conf.\ Ser.\ 120 (2008) 052057. 
\bibitem{Scielzo} N.D.~Scielzo et al., Phys.\ Rev.\ C 80 (2009) 025501.
\bibitem{EC} W.~Bambynek et al., Rev.\ Mod.\ Phys.\ Vol.49, No.1 (1997). 
\bibitem{Qino08} C.~Arnaboldi et al., Phys.\ Rev.\ C 78 (2008) 035502.
\bibitem{TOI} http://nucleardata.nuclear.lu.se/nucleardata/toi/ .  
\bibitem{Qino98} A.~Alessandrello et al., Nucl.\ Instrum. and Meth.\ B 142 (1998) 163.
\bibitem{Arn02} C.~Arnaboldi et al., IEEE Trans.\ Nucl.\ Sci. 49 (2002) 2440. 
\bibitem{Qino04} C.~Arnaboldi et al., Nucl.\ Instr.\ Meth. A 518 (2004) 775.
\bibitem{Arn03} C.~Arnaboldi et al., IEEE Trans.\ Nucl.\ Sci. 50 (2003) 979.
\bibitem{PDG} C.~Amsler et al. (Particle Data Group), Physics Letters B667, 1 (2008).
\bibitem{CUORE} R.~Ardito et al. (CUORE Collaboration), arXiv:hep-ex/0501010. 

\end{thebibliography}
\end{document}